\documentclass{ws-procs975x65}

\begin{document}

\title{Diffractive Microlensing: A New Probe of the Local Universe}
\author{Jeremy S. Heyl}
\address{Department of Physics and Astronomy, University of British
  Columbia, Vancouver BC Canada}
\begin{abstract}
  Diffraction is important when nearby substellar objects
  gravitationally lens distant stars. If the wavelength of the
  observation is comparable to the Schwarzschild radius of lensing
  object, diffraction leaves an observable imprint on the lensing
  signature. The SKA may have sufficient sensitivity to detect the
  typical sources, giant stars in the bulge. The diffractive
  signatures in a lensing event break the degeneracies between the
  mass of the lens, its distance and proper motion.
\end{abstract}

\section{Preliminaries}

As light passes an object, it diffracts around it and spreads out by
an angle $\sim \lambda/L$, so if the angle that the object subtends,
$L/d$, is less than this, the object ``disappears''.  Furthermore, if
the object has mass, light that passes close-by is delayed by
travelling through its potential well.  The figure of merit for
whether diffraction is important in gravitational lensing is $f=4 \pi
R_S/\lambda$ where $R_S=2G M/c^2$ is the Schwarzschild radius of the
lens.  For values of $f$ much less than unity, diffraction dominates
and the peak magnification is greatly diminished,  For large values of
$f$ the diffraction fringes are closely spaced, so either the finite
size of the source or the finite bandwidth of the observations can
easily smear them out yielding the magnification that one expects from
geometric optics.

The path length from the star at $\vec v$ to the telescope through the point
$\vec u$ in the lens plane is
\begin{equation}
\Delta  l = \frac{\left | \vec u - \vec v \right |^2}{2} =
\frac{u^2}{2} - u v \cos \varphi + \frac{v^2}{2} - f \ln u.
\end{equation}
The magnification of the star is related to the sum of all the phases from 
the light taking various paths,
\begin{equation}
\mu_\omega =
\left | \int_{u_d}^\infty du  u^{1-if} e^{iu^2/2} 
\int_0^{2\pi} d\varphi  e^{-iuv\cos \varphi} \right |^2.
\end{equation}
The Schwarzschild radius of the Earth is about one centimeter.
Beyond $10^4$~AU lensing dominates over occultation.
Therefore, let's take $u_d \rightarrow 0$. If we take the integral over the amplitudes to be $I_\omega$
($\mu_\omega=|I_\omega|^2$),
\begin{equation}
I_\omega = e^{\pi f/4} e^{i \left ( \pi - f \ln 2\right)/2}
  \Gamma \left (1 - i \frac{f}{2}\right ) 
{}_1F_1 \left ( 1 - i \frac{f}{2}; 1 ; -i\frac{v^2}{2} \right ).
\end{equation}
The magnification increases with the value of $f$ with the peak
magnification given by ${\pi f}/(1-e^{-\pi f})$. 
\begin{figure}
\epsfig{file=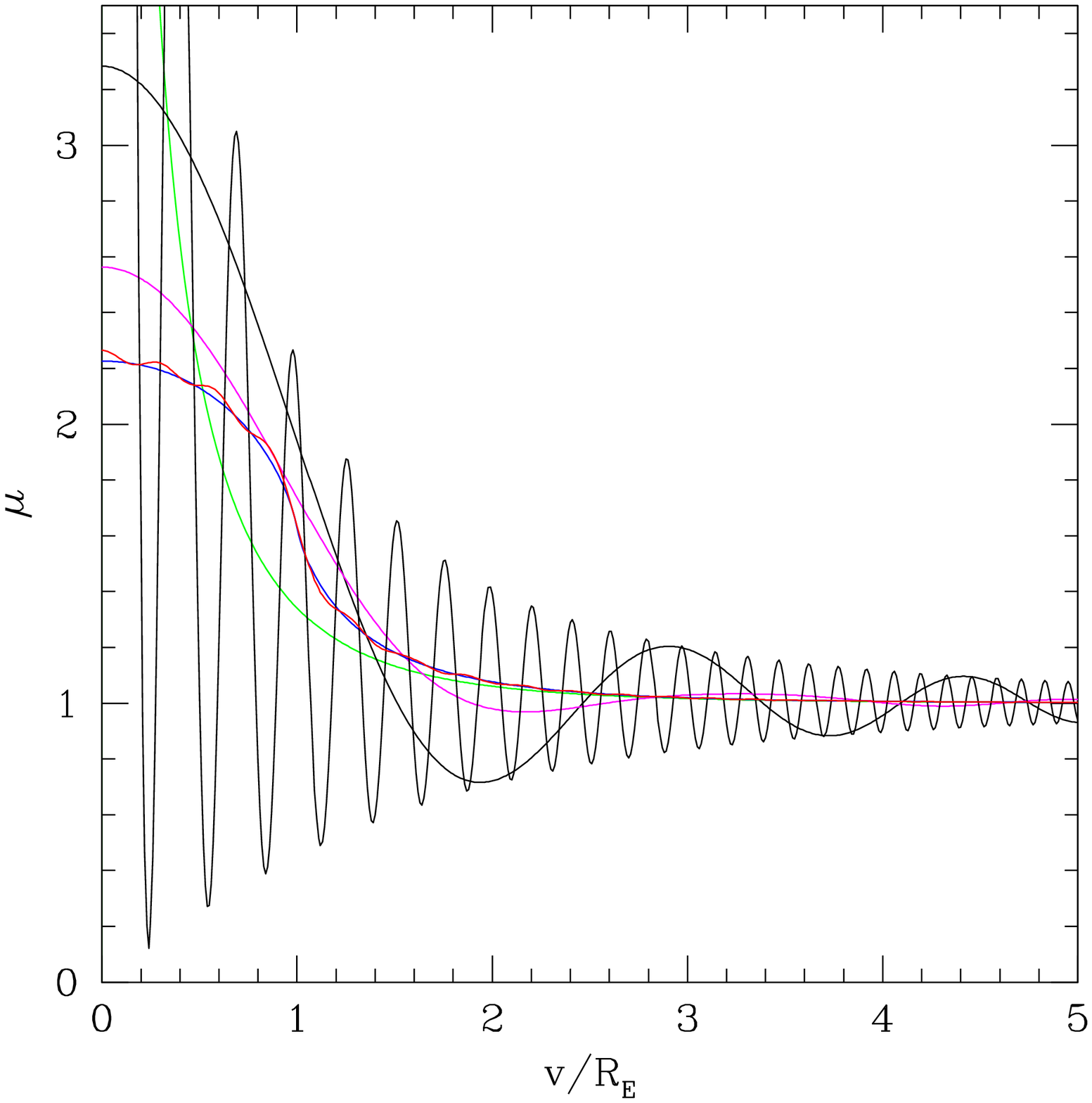,width=0.5\linewidth,clip=} 
\epsfig{file=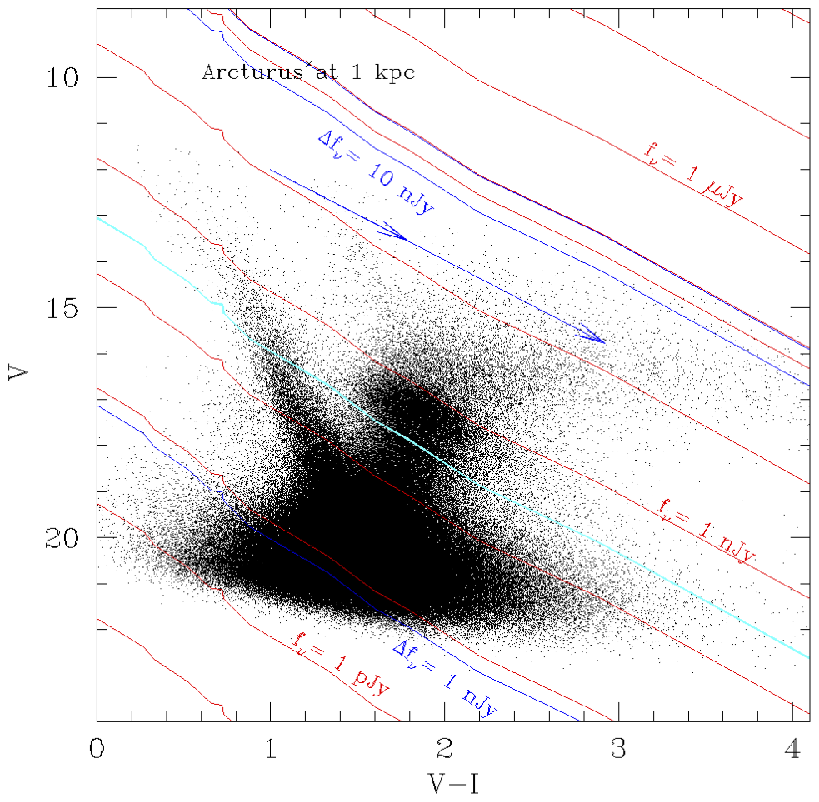,width=0.48\linewidth,clip=} 
\caption{Left panel: 
The values of the magnification $\mu$ as a function of the
  distance between the source and the centre of the lens in units of
  the Einstein radius, $v/R_E$.  The result for $f=1$ is the slowly
  varying sinusoidal curve and $f=10$ is the more rapidly varying
  one.  The magnification from geometric optics is plotted in green.
  Notice for $f=10$ there are about three peaks over a length of one
  Einstein radius.  The other colours assume that the angular radius of the
  source equals the Einstein radius.  Blue is the geometric optics
  result, red is for $f=10$ and magenta is $f=1$
Right panel: The blue lines give the expected oscillation at 10~GHz induced by an
Earth-mass planet at one parsec microlensing an OGLE source.
About two percent of all OGLE-II bulge sources would have a detectable 
diffractive microlensing signal with the SKA.}
\label{fig:finite_sources}
\end{figure}
The left panel of figure~\ref{fig:finite_sources} depicts the
magnification as function of the impact parameter of the source
relative to the lens in units of the Einstein radius and the value of
$f$.  Averaging over bandwidth or over a finite source brings the
oscillations toward the geometric.  However, if $f<1$ there is little
magnification.  For large values of $f$, the size of the oscillation
scales as $(f\theta_S/\theta_E)^{-3/2}.$ The right panel of
figure~\ref{fig:finite_sources} gives a estimate of the radio flux
density (red) and amplitude of its oscillation (blue) at 10~GHz for
stars in the bulge field 1 of the OGLE-II survey (Szymanski 2005;
Udalski, Kubiak \& Szymanski 1997).  The fluxes are normalized using
the values for Arcturus at 1~kpc (consult Heyl (2010b) for details).

\section{Astrometry}

\begin{figure}
\epsfig{file=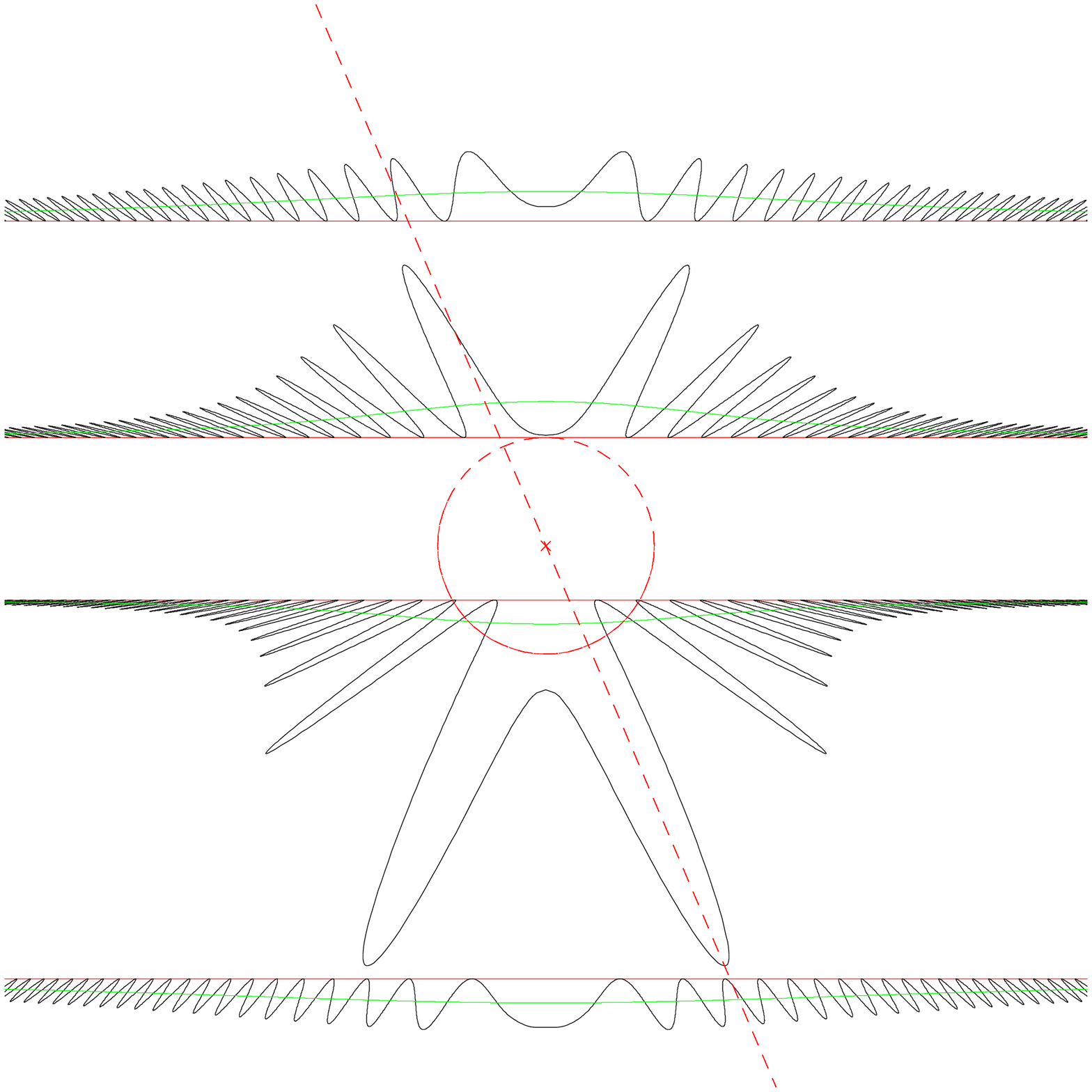,width=0.49\linewidth,clip=} 
\epsfig{file=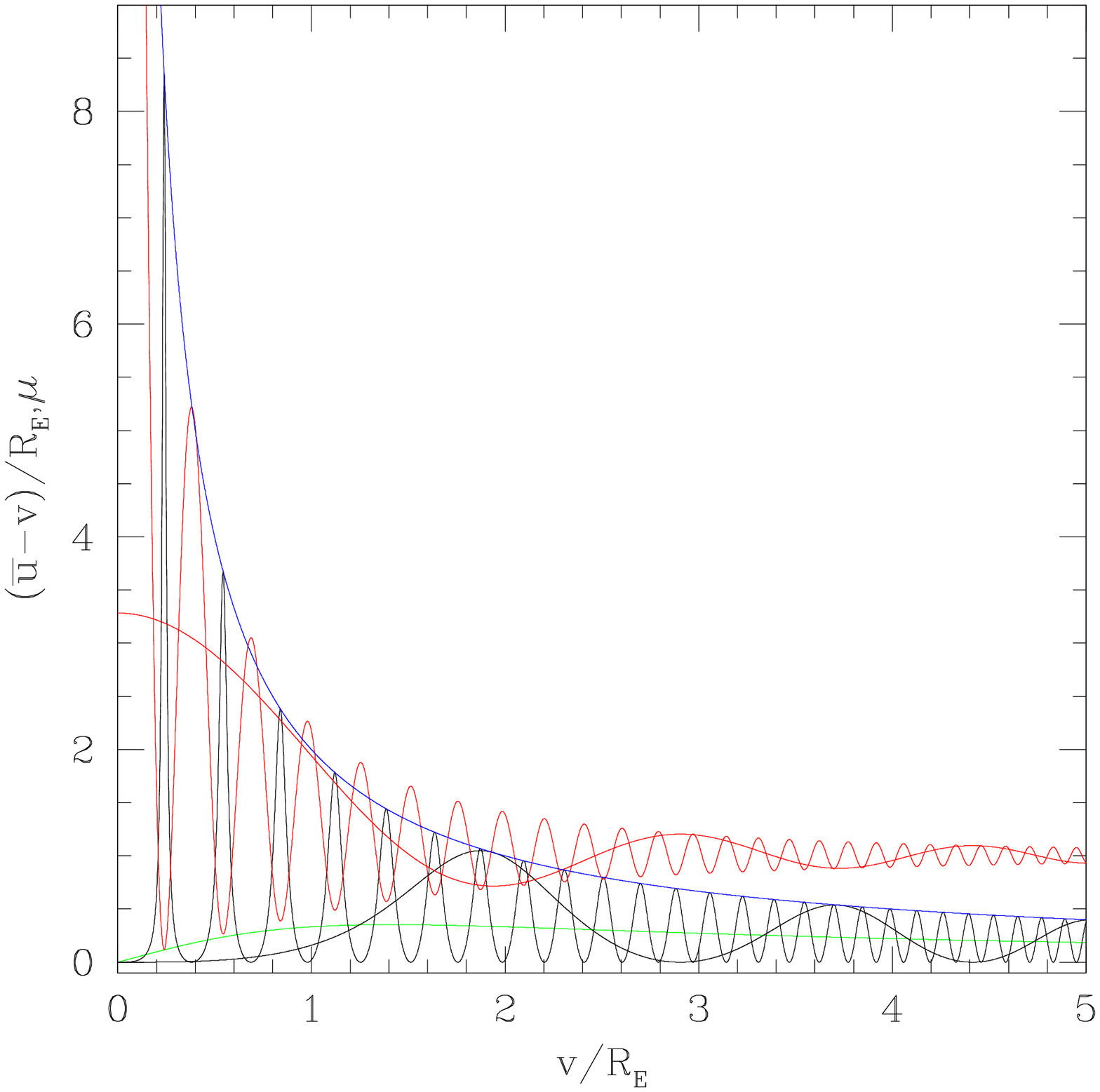,width=0.49\linewidth,clip=} 

\caption{The left panel gives the trajectory of the centroid: black
  denotes $f=10$, green denotes the geometric limit and red is the
  unlensed trajectory. The dashed circle denotes the Einstein radius
  of the lens. The right panel gives the total magnification in red
  and the motion of the image in black for $f=10$ and $f=1$.}
\label{fig:astrometry} 
\end{figure}

There is a second complementary signature (Heyl 2010c).  As the lens
passes in front of the source, its position on the sky appears to
move.  The key idea to figure how where the image appears is that the
gradient of the phase of the wave as measured at the telescope points
toward the image.  Furthermore, the situation is symmetric about the
lens, so moving the telescope to the right is like moving the source
to the left, and we can take the gradient of the phase with respect to
the position of the source instead at the telescope, yielding
the centroid of the image
\begin{equation}
{\bar u} = -\Im \frac{\partial \ln I_\omega}{\partial v} 
 = v \Re \left [   \left ( 1 - i \frac{f}{2} \right )
  \frac{{}_1F_1 \left ( 2 - i \frac{f}{2}; 2 ; -i\frac{v^2}{2} \right
    ) }{{}_1F_1 \left ( 1 - i \frac{f}{2}; 1 ; -i\frac{v^2}{2} \right
    ) } \right ]
\end{equation}
Figure~\ref{fig:astrometry} shows that lensing moves the centroid
radially outward: the closer the passage, the greater the deviation.
Minima in the displacement correspond to maxima of the magnification,
and the displacement reaches the blue curve $2 R_E^2/v$ at minima of the
magnification.

\section{Conclusions}

For distant asteroids $\lambda/L\sim L/d$ (Heyl 2010a).  If they are
massive, lensing is also important.  If lensing is weak, the Fresnel
pattern gets lensed.  If lensing is strong, things are more complex.
The SKA could detect oscillations from Earth-mass lenses as far as 10~pc
against bulge giants.

\section*{Bibliography}
Heyl, J., 2010a, {\em MNRAS}, {\bf 402}, L39-L43 (arXiv:0910.3922). \\
---, 2010b, {\em MNRAS}, {\bf 411}, 1780-1786 (arXiv:1002.3007). \\
---, 2010c, {\em MNRAS}, {\bf 411}, 1787-1791 (arXiv:1003.0250).
 \\
Schneider, P., Ehlers, J., Falco, E. 1992, {\em Gravitational Lenses},
Springer: Berlin.\\
{Szymanski} M.~K.,  2005, {\em Acta Astron.}, 55, 43.\\
{Udalski} A.,  {Kubiak} M.,    {Szymanski} M.,  1997, {\em Acta Astron.}, 47, 319.
\end{document}